\documentstyle[aps,prb,epsf,floats,twocolumn]{revtex}

\begin{document}
\draft
\twocolumn[\hsize\textwidth\columnwidth\hsize\csname @twocolumnfalse\endcsname

\title{Monte Carlo study of a two-dimensional quantum ferromagnet}
\author{Patrik Henelius$^a$, Anders W. Sandvik$^b$, Carsten Timm$^c$ and 
S. M. Girvin$^d$}
\address{$^a$ National High Magnetic Field Laboratory, Tallahassee, FL 32310 \\
$^b$ Department of Physics, University of Illinois at Urbana-Champaign,
Urbana, Illinois  61801\\
$^c$ Institut f\"ur Theoretische Physik, Freie Universit\"at Berlin, 
Arnimallee 14, D-14195 Berlin, Germany\\
$^d$ Department of Physics,Indiana University, Bloomington, Indiana 47405 }

\date{\today}  
\maketitle

\begin{abstract}
We present quantum Monte Carlo results for the field and temperature
dependence of the magnetization and the spin-lattice relaxation rate
$1/T_1$ of a two-dimensional $S=1/2$ quantum Heisenberg
ferromagnet. The Monte Carlo method, which yields results free of
systematic errors, is described in detail. The high accuracy of the
calculated magnetization allows for stringent tests of recent
approximate analytical calculations. We also compare our results with
recent experimental data for a $\nu=1$ quantum Hall ferromagnet, which
is expected to be well described by the Heisenberg model. The dynamic
response function needed to extract $1/T_1$ is obtained using
maximum-entropy analytic continuation of the corresponding
imaginary-time dependent correlation function. We discuss the
reliability of this approach.
\end{abstract}
\vskip2mm]

\section{INTRODUCTION}
The magnetization of a two-dimensional electron gas in the quantum
Hall regime has recently been measured.~\cite{Barr,Manf} The
realization that this system represents a novel itinerant ferromagnet
has motivated several theoretical studies.~\cite{Sond,Cote,Girv,Ezaw} The
system at filling factor $\nu=1$ should be well described by a
two-dimensional Heisenberg ferromagnet~\cite{Read}, and in a recent
publication we presented analytical Schwinger boson and numerical
Monte Carlo results for this model.~\cite{Timm} The ferromagnetic
Heisenberg model has been much less studied than the antiferromagnetic
model, probably because the ground state and lowest excitations are
known exactly. At finite temperatures there is, however, no exact analytic
solution, and in our first paper we gave a detailed discussion of the
calculation of the leading $1/N$ correction to the Schwinger boson
mean-field theory. In this second paper we want to give a detailed
description of the Monte Carlo calculation and present results for the
NMR relaxation rate $1/T_1$. The Monte Carlo technique used is highly
efficient and suffers from none of the common systematic errors. The
magnetization results are accurate to a relative statistical error of
about $10^{-4}$ and to this accuracy they are also free of finite size
corrections. This allows for stringent tests of the analytical
results. In order to estimate $1/T_1$ we calculate spin correlation
functions in imaginary time and continue these to real frequency using
the maximum entropy method. We discuss how the results of this
approach depend on whether real- or momentum space correlation
functions are analytically continued. 

We want to emphasize an important technical detail that makes the
sampling particularly efficient: the external field is chosen in the
$x$-direction, perpendicular to the $z$-direction, in which the basis
states are expressed. This automatically causes the simulation to
become free of systematic errors, even though only local updates are
used. Furthermore, it enables easy access to observables involving
both diagonal and off-diagonal operators.

In Sec.~II the stochastic series expansion Monte Carlo technique is 
briefly reviewed and applied to the two-dimensional ferromagnet.
The maximum entropy technique used to numerically perform the analytic 
continuation to real frequency is described in Sec.~III. Results for 
the field and temperature dependence of the magnetization are given in 
Sec~IV.~A, and the NMR relaxation rate results are presented in Sec~IV.~B. 
Our results and conclusions are summarized in Sec.~V.

\section{QUANTUM MONTE CARLO}
\subsection{Introduction}
Over the last few decades, several methods for quantum Monte Carlo
calculations have been developed. The most common methods
stochastically sample the configurations of the world-line path
integral of the system.~\cite{Revi} These methods traditionally suffer
from several systematic errors.~\cite{Prok} The Trotter discretization
of the imaginary time introduces a systematic error that decreases
with decreasing ``time slice'' width $\Delta\tau$. In principle exact
results are obtained by scaling to $\Delta\tau \to 0$. However, with
standard techniques,~\cite{Hirs} the efficiency of the simulation
decreases rapidly as $\Delta\tau$ is decreased,~\cite{Prok} and it may
therefore be difficult to obtain results completely free of systematic
errors. Each configuration can furthermore be labeled by a topological
``winding number''. The acceptance rate for moving from one winding
number sector to another gets extremely small as the system size
increases, thereby making the simulation non-ergodic.~\cite{Hene} In
addition, at low temperatures it can be very hard to change the number
of particles in fermion or boson simulations, or the total
magnetization for spin systems, hence restricting simulations to a
canonical ensemble.

Recently there has been much progress in resolving these technical
problems, making it much easier to obtain results that are exact
within statistical error bars. Non-local  ``loop cluster'' moves 
(analogous to cluster updates for classical systems~\cite{Swen}) have
been developed for several models~\cite{Ever,Weis} and can
significantly reduce the autocorrelation times of the simulations.
They enable sampling of all winding number sectors and all magnetizations.
In addition these moves have been formulated in continuous imaginary 
time,~\cite{Bear} hence eliminating the Trotter error and making the 
simulation completely free of systematic errors. A related method was
recently formulated starting from the standard perturbation expansion 
in the interaction representation.~\cite{Prok} Within this formulation, 
new local moves that share some of the advantages of the loop
algorithms was introduced.

A different approach to quantum Monte Carlo simulation is the so
called stochastic series expansion (SSE) method ~\cite{Sand,Loop} (a
generalization of Handscomb's method~\cite{Hand} to a much larger
class of models) which does not use the Trotter decomposition as a
starting point, but instead samples matrix elements of the exact
Taylor expansion of the density matrix $e^{-\beta H}$. The Trotter
error is thus automatically avoided. There are strong formal
relationships between the SSE formulation and the continuous-time path
integral, which have been discussed elsewhere~\cite{Sand2}. We will
here demonstrate that when applying the SSE method to a ferromagnet in
a magnetic field, purely local moves are sufficient to make the
calculation ergodic in the grand canonical ensemble.

Below we will first give a short general overview of the SEE method, 
and thereafter give a detailed description of the application to the
Heisenberg ferromagnet in a magnetic field. 

\subsection{Stochastic Series Expansion}
The thermodynamic expectation value for an operator
$A$, at inverse temperature $\beta$, is
\begin{equation}
\langle  A \rangle ={ \mbox{Tr} \lbrace A e^{\beta  H}\rbrace
   \over \mbox{Tr} \lbrace e^{\beta  H} \rbrace }= {1 \over Z} 
   \mbox{Tr} \lbrace A e^{-\beta  H}\rbrace.
\label{eq:exp}
\end{equation}
Assume that the Hamiltonian consists of $M$ terms that need not  
commute, and may be of any order in field operators:
\begin{equation}  
H= -\sum_{i=1}^M H_i.
\label{hsum}
\end{equation}
A minus sign has been pulled out of the sum for convenience.
The density matrix $\exp(-\beta H)$ in the expectation value 
Eq.\ (\ref{eq:exp}) is Taylor-expanded,
\begin{equation} 
\langle  A \rangle= {1\over Z}\sum_{n=0}^{\infty}\sum_{S_n}
 {\beta^n \over n!}\mbox{Tr}\lbrace  A \prod_{i=1}^n  H_{k_i} 
  \rbrace ,
\end{equation}
where $S_n$ denotes an index sequence $k_1,\ldots,k_n$ with $1 \le k_i \le
M $, and $\prod_{i=1}^n H_{k_i}$ is an operator string of length $n$.
If the above trace can be analytically calculated the expectation value
can be calculated by importance sampling in the space of index sequences.
In the original Handscomb's method~\cite{Hand}
a spin-1/2 system is considered, for which the trace can be evaluated 
analytically. In the more general SSE method~\cite{Sand}
a complete set of states $\{\vert \alpha\rangle\}$ is introduced to 
calculate the trace, and hence a larger class of problems can be treated:
\begin{equation}
\langle  A \rangle= {1\over Z}\sum_{\alpha}\sum_{n=0}^{\infty}
\sum_{S_n} {\beta^n \over n!}\langle \alpha \vert  A
\prod_{i=1}^n  H_{k_i} \vert \alpha \rangle. 
\label{eq:exp2}
\end{equation}
In order to calculate the expectation value of the operator $A$
we assume that a function $A(\alpha,S_n)$ exists such that 
Eq.\ (\ref{eq:exp2}) can be re-written as:
\begin{eqnarray}
\langle  A \rangle && = {\sum_{\alpha}\sum_{n=0}^{\infty}\sum_{S_n} 
A(\alpha,S_n) W(\alpha,S_n)\over \sum_{\alpha}\sum_{n=0}^{\infty}\sum_{S_n} 
W(\alpha,S_n) } \nonumber \\
&& =\langle  A(\alpha,S_n)\rangle_W,
\label{expval}
\end{eqnarray}
where the weight function $W(\alpha,S_n)$ is given by
\begin{equation} 
W(\alpha,S_n) = {\beta^n \over n!} \langle\alpha| \prod_{i=1}^n  H_{k_i}
|\alpha\rangle.
\label{wn}
\end{equation}
We will assume that $W(\alpha,S_n)$ is positive definite, which normally 
is the condition for a stochastic evaluation of (\ref{expval}) to be feasible.
One can then carry out a random walk satisfying the detailed-balance principle
in the space $ \lbrace \vert \alpha \rangle \otimes \lbrace S_n, 
n=0,\ldots, \infty\rbrace \rbrace $, using $W$ as a relative probability.
One important condition for such a procedure to be feasible in practice
is that the operators $\hat H_i$ in Eq.~(\ref{hsum}) must be ``non-branching'',
i.e., the application of any $H_i$ to a single basis vector should
yield a single basis vector (not a linear combination of them),
\begin{equation}
 H_i\vert\alpha\rangle=h(i,\alpha)\vert\alpha'\rangle , \qquad
\vert\alpha\rangle,\vert\alpha'\rangle \in \lbrace\vert\alpha\rangle
\rbrace,
\end{equation}
so that the weight factor (\ref{wn}) can be easily calculated. A scheme 
can then be constructed in which first an arbitrary allowed operator 
string and state are chosen. Thereafter relatively simple updating 
steps (``moves'') are performed that change the number of operators
$(n)$ in the string, the individual operators within the string 
(thereby changing $S_n$), and the state $\vert \alpha \rangle$. The 
acceptance probabilities for the moves are chosen so that the 
detailed-balance principle is satisfied, using, e.g., the standard
Metropolis or heat-bath algorithm.

\begin{figure}
\centering
\epsfxsize=8cm
\leavevmode
\epsffile{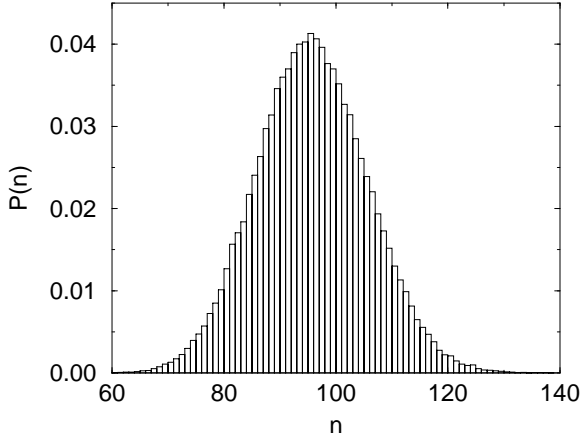}
\vskip0.4cm
\caption{Distribution of the order of the Taylor expansion for a 
$4 \times 4$ Heisenberg ferromagnet at  temperature $T/J=2.0$
and magnetic field $h/J=1.0$.}
\label{fig:nophist}
\end{figure}

The Taylor expansion may appear to be a high-temperature expansion, 
but in principle terms  up to $n=\infty$ are included, and the expansion 
is equally valid at any temperature. For a finite-size
system at a finite temperature only terms of finite length contribute 
significantly to the trace, and importance sampling includes all relevant 
terms.  This can be compared to the Taylor expansion of the exponential
of a simple number, where the factorial  $n!$  in the denominator eventually 
wins over the numerator. An actual distribution for the order of the series in
a typical simulation appears to be close to a normal distribution; 
see Fig.~\ref{fig:nophist}. As will be derived below, the average length of 
the operator string, $\langle n\rangle$, 
equals $\beta E$, where $\beta$ is the inverse
temperature and $E$ is the total energy. At low temperatures we thus see
that the average operator-string length is inversely proportional to the
temperature and proportional to the number of sites. The computational
time required for one MC step is proportional to the inverse
temperature times the system size. The computational cost to 
achieve a given accuracy is, however, often offset by the fact that
as the length of the operator-string is increased it also contains
more information about the system. This will become clear when we
discuss how to measure various expectation values below.

Next, we need to find the function $A(\alpha,S_n)$ for different cases
of interest. First we look at some important features of the states
and introduce some notation. It is convenient to introduce states 
$\vert\alpha(p)\rangle$ that are obtained by propagating $\vert\alpha\rangle$
with $p$ ($p=1,\ldots,n$) of the operators in Eq.~(\ref{wn});
\begin{equation}
\vert\alpha(p)\rangle=r\prod_{i=1}^p H_{k_i}\vert\alpha\rangle ,
\end{equation}
where $r$ is a normalization factor. With this definition 
$\vert\alpha(0)\rangle=\vert\alpha\rangle $. Using this notation 
the matrix element in the weight function is
\begin{eqnarray}
\langle\alpha\vert && \prod_{i=1}^n H_{k_i}\vert\alpha\rangle = \\
&& \left\{ \begin{array}
{l@{}l}
\prod_{i=1}^n h\lbrack k_i,\alpha(i-1)\rbrack ,&\quad \hbox{if } \vert
\alpha(n)\rangle=\vert\alpha(0)\rangle \\
0,& \quad\hbox{otherwise}.
\end{array} \right. \nonumber
\end{eqnarray}
If the index sequence $S_n$ is cyclically permuted $p$ times 
we obtain $k_{p+1},\ldots ,k_n, k_1,\ldots, k_p$, which is denoted
$S_n(p)$. Since $W(\alpha,S_n)=W(\alpha(p),S_n(p))$ it follows
that
\begin{equation}
\langle  A\rangle = \Bigl\langle {1\over n+1} \sum_{p=0}^n A\lbrack
\alpha (p), S_n(p)\rbrack\Bigr\rangle_W.
\label{eq:ave}
\end{equation}
In the simplest case where $A$ is diagonal, $ A\vert\alpha\rangle=
a(\alpha) \vert \alpha\rangle$, we find that $A(\alpha,S_n)=a(\alpha)$
and
\begin{equation}
\langle A\rangle =  \Bigl\langle {1\over n+1} \sum_{p=0}^{n}
a\lbrack \alpha(p)\rbrack\Bigr\rangle_W.
\label{diaexp}
\end{equation}

Next we treat the case where $ A= H_m$, where $H_m$ is one of the
terms in the Hamiltonian, then
\begin{equation}
\langle  A \rangle =\langle  H_m \rangle = {1\over
Z}\sum_{\alpha}\sum_{n=0}^{\infty} \sum_{S_n} {\beta^n \over
n!}\langle \alpha \vert  H_m \prod_{i=1}^n  H_{k_i} \vert \alpha
\rangle , 
\end{equation}
and for each index sequence $S_n$ for the state $\vert \alpha\rangle $,
there is a sequence $S_{n+1}$, with $H_m$ as the last element. Defining
\begin{equation}
A(\alpha ,S_n)=
\left\{ \begin{array}
{l@{}l}
 {n\over \beta},  &\quad k_n=m \\
                0, &\quad k_n\not= m,
\end{array} \right.
\end{equation}
and considering Eq.~({\ref{eq:ave}), the expectation value is obtained by 
counting the number $N(m)$ of $H_m$ operators in the sequence,
\begin{equation}
\langle  H_m\rangle\ = {1\over \beta}\Bigl\langle 
N(m)\Bigr\rangle_W.
\label{exphm}
\end{equation}

The energy is the negative sum of all operators $H_m$, and it therefore follows
that
\begin{equation}
 E=-{1\over\beta}\langle n\rangle_W,
\label{energy}
\end{equation}
and we see that the energy is proportional to the average length of
the operator sequence. One may argue that this is a strange formula;
it seems that one should be able to decrease the average string length
by adding a positive constant to the Hamiltonian (thereby increasing
the energy). One should even be able to set the
energy and average string length to zero! The solution to this
paradox lies in the infamous sign problem. Adding a (sufficiently large)
positive constant to the Hamiltonian will make it impossible to keep 
the weight function positive, and hence Eq.\ (\ref{energy}) is no 
longer valid. In fact, for most models, a {\it negative} constant has
to be added to the Hamiltonian in order to make $W(\alpha,S_n)$
positive definite. Eq.~(\ref{energy}) then gives the energy including 
these constants.

Other expectation values, such as products of imaginary-time dependent
operators and static susceptibilities can also be easily evaluated with
the SSE method. We refer to previous work for the expressions for
these more complicated expectation values.~\cite{Sand}

In order to make the simulation code efficient one assumes that only 
operator strings with a length shorter than $L$ contribute to the
trace. This is not necessary, but by automatically adjusting $L$ so that 
the simulation will not reach strings longer than $L$ in any reasonable 
simulation time, only an exponentially small, completely undetectable,
error is made. With the length of the operator string limited to 
$L$, $(n-L)$ identity operators can introduced in an operator string 
of length $n<L$ to make the string length fixed. For every original operator 
string of length $n$ there are then $L\choose n$ strings of length $L$, 
corresponding to all possible insertions of the $(L-n)$ unit operators.
The inverse of this factor is included in the Taylor series expansion, 
where the summation now is over all index sequences of length $L$, denoted
$S_L$:
\begin{equation}
\langle  A \rangle= {1\over Z}\sum_{\alpha}
\sum_{S_L} {(-\beta)^n \lbrack L-n \rbrack !\over L!}\langle \alpha
\vert  A \prod_{i=1}^L H_{k_i} \vert \alpha \rangle. 
\end{equation}
During the simulation the number of these extra unit operators will
fluctuate, and hence the previous sum over $n$ is implied. The advantage
with fixing the length in this manner and introducing extra unit operators
is that all moves can be defined in solely terms of exchanges, i.e. a
set of one or several operators is exchanged for another set of the
same number of operators. This simplifies the construction of an 
updating scheme that satisfies detailed balance.

\subsection{Application to the Heisenberg ferromagnet}
In this Section we give a detailed description of how
the SSE method is applied to the Heisenberg ferromagnet. 
The Hamiltonian of this model including a magnetic field is given by
\begin{equation}
H = -J \sum_{i,\delta (i)} \vec S_i\cdot \vec S_{\delta (i)} - h\sum_i S^x(i),
\end{equation}
where the coupling constant $J>0$, $h$ denotes the magnetic field strength
and $\delta (i)$ denote the nearest neighbors of site $i$ (we count
each interacting spin pair only once). For reasons that will
be explained later, the magnetic field is chosen in the $x$ direction. 
The rectangular lattice has $n=l_1\times l_2$ sites, where $l_1$ and 
$l_2$ are the linear dimensions of the rectangular lattice. Throughout 
this work we use periodic boundary conditions, and there are therefore 
$m=n$ bonds if $l_1=1$, corresponding to the one-dimensional lattice, 
and $m=2n$ bonds if $l_1>1$, corresponding to the two-dimensional case. 

For the purpose of the SSE updating scheme we introduce the following 
six operators:
\begin{eqnarray}
\nonumber H_{0,0}&=&I\\
\nonumber H_{1,b}&=&I_b\\
 H_{2,i}&=&I_i\\
\nonumber H_{3,b}&=&2(S_{i(b)}^zS_{j(b)}^z + {\hbox{$1\over 4$}} I_b)\\
\nonumber H_{4,i}&=&S_i^++S_i^-\\
\nonumber H_{5,b}&=&S_{j(b)}^+S_{k(b)}^-+S_{j(b)}^-S_{k(b)}^+,
\end{eqnarray}
where $i$ denotes a lattice site, $b$ a bond, $i(b)$ and $j(b)$ are
the sites connected by bond $b$, and $I$ is an identity operator. 
The unit operators $I_b$ and $I_i$, labeled by a site or bond, are 
introduced to simplify the updating scheme. An algorithm without 
these could also be formulated. 

The Hamiltonian can then be written as
\begin{eqnarray}
\nonumber H&=&-{J\over 2}\sum_{b=1}^m(H_{1,b}+H_{3,b}+H_{5,b})\\
 &-&{h\over 2}\sum_{i=1}^n(H_{2,i}+H_{4,i}) + {3Jm\over 4}+{hn
 \over 2}.
\end{eqnarray}

Note that the operator $H_{0,0}=I$ is not considered a term of the
Hamiltonian; it is the unit operator employed to augment the operator
strings to the fixed length $L$, as discussed in the previous
Section. The constant in the definition of the type-3 operator has
been introduced in order to make all its matrix elements positive (or
zero), thereby making the weight function positive definite. The
introduced constants only shift the energy (and therefore also the
mean length of the operator string).

According to an ergodic updating scheme, the six different kinds of 
operators in the operator sequence are interchanged in such a way 
that any contributing operator string and basis state can be 
generated by a series of updates. Before going into the details 
of these procedures, we will specify our basis states and the weight 
function, and introduce some notation.

We will work in the $S^z$ basis $|S_1^z,S_2^z,\ldots,S_n^z\rangle$,
and the non-branching property is satisfied since
\begin{eqnarray}
\nonumber H_3|\uparrow\downarrow\rangle&=& H_3|\downarrow\uparrow\rangle=0\\
\nonumber H_3|\uparrow\uparrow\rangle&=&|\uparrow\uparrow\rangle\\
\nonumber H_3|\downarrow\downarrow\rangle&=&|\downarrow\downarrow\rangle\\
 H_4|\uparrow\rangle&=&|\downarrow\rangle\\
\nonumber H_4|\downarrow\rangle&=&|\uparrow\rangle\\
\nonumber H_5|\uparrow\uparrow\rangle&=& H_5|\downarrow\downarrow\rangle=0\\
\nonumber H_5|\downarrow\uparrow\rangle&=&|\uparrow\downarrow\rangle\\
\nonumber H_5|\uparrow\downarrow\rangle&=&|\downarrow\uparrow\rangle .
\label{hdefs}
\end{eqnarray}

When non-zero, i.e., if $|\alpha (L) \rangle  = |\alpha (0) \rangle$,
the weight function is
\begin{equation}
W(\alpha,S_L)=\beta^n(L-n)! \left  ({J\over 2} \right )^{n_1+n_3+n_5}
\left ({h \over 2}\right )^{n_2+n_4} , 
\label{wl}
\end{equation}
where $n_k$ is the total number of $H_{k,i}$ ($i=1,\ldots,m$) operators 
in the operator string. The weight is always positive because all
the matrix elements of the operators defined in Eq.~(\ref{hdefs}) are
positive or zero. For convenience we now use a two-index notation 
for the operator-index sequence $S_L$; 
\begin{equation}
S_L={k_1 \choose j_1}_1{k_2 \choose j_2}_2\cdots {k_L \choose j_L}_L,
\end{equation}
where $k$ indicates the kind of operator $(k\in \{0,\cdots,5\})$ and
$j$ indicates what site or bond the operator acts on.

Next we introduce a random walk that satisfies the detailed-balance
principle and covers the space $ \lbrace \vert \alpha \rangle \otimes 
\lbrace S_n, n=0,\ldots, L\rbrace \rbrace $. A simulation is started
from a random initial state $|\alpha(0)\rangle$ and an initial sequence 
of operators ${0 \choose 0}_1{0 \choose 0}_2\cdots {0 \choose 0}_L$. 
We define six fundamental moves that together ensure that
we cover the full configuration space as long as the magnetic field 
$h$ is nonzero. In the absence of a magnetic field, additional moves 
have to be made, and these will be described later. The moves have to 
be carefully defined so that the detailed-balance principle is satisfied, 
generation of zero-weight configurations is not attempted, and so that
they can be rapidly executed.

To visualize the seemingly abstract SSE space, a typical configuration for
a one-dimensional four-site system is shown in Fig. \ref{fig:sse}. 
In the figure we can also see the close resemblance between the SSE 
method and the standard Euclidean path integral formulations. This 
relation has been explored in detail.~\cite{Sand2} Note that only the 
first state $|\alpha (0)\rangle$ and the operator string have to be 
stored in memory. All propagated states can be generated sequentially
as needed, and the memory requirements for the method are therefore 
quite modest.

\begin{figure}
\centering
\epsfxsize=4cm
\leavevmode
\epsffile{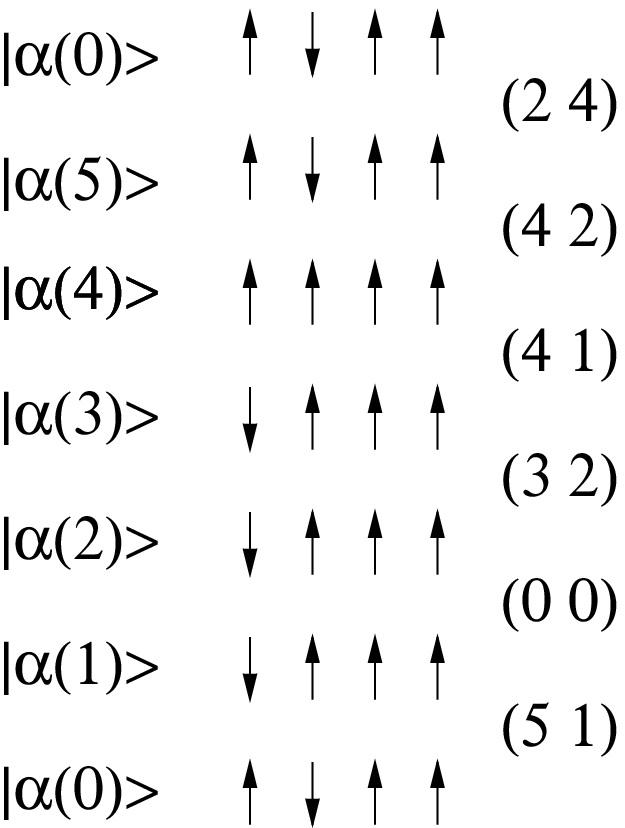}
\vskip0.4cm
\caption{Typical SSE configuration for a four-site system. The propagated 
states are shown on the left and the operator string to the right.
The first operator-index denotes the type of operator ($1-6$) and the 
second is the bond on which it operates ($1-m$), with $m=4$ in this case.}
\label{fig:sse}
\end{figure}

The first kind of move changes the number of nonzero
operators in the string by introducing the unit operator
of type 1. Going through the operator string from the
first to the last operator, attempts are made to exchange every 
${0\choose 0}_p$ operator for a ${1\choose b}_p$ operator (with $b$ 
chosen randomly from $\{1,\ldots,m\}$) and every ${1\choose b}_p$ operator 
for a ${0\choose 0}_p$ operator. In an accepted move $n$ and $n_1$ in 
Eq.~(\ref{wl}) changes by $\pm 1$. The detailed-balance principle then
requires that the probabilities $P\lbrack {0\choose 0}_p \leftrightarrow 
{1\choose b}_p\rbrack$ of carrying out the replacements in the
two different directions satisfy
\begin{equation}
\nonumber {P\lbrack {0\choose 0}_p \to {1\choose b}_p\rbrack \over
P\lbrack {1\choose b}_p \to {0\choose 0}_p\rbrack} = 
{W(\alpha,S_{n+1})p[{1\choose b}_p \to {0\choose 0}_p] \over 
W(\alpha,S_{n}) p[{0\choose 0}_p \to {1\choose b}_p]}
\end{equation} 
where $m$ again is the number of bonds and 
$p\lbrack {0\choose 0}_p \leftrightarrow {1\choose b}_p\rbrack$ 
denotes the {\it a priori}
probability of the move being carried out, i.e., the probability before any
accept/reject probability has been assigned. If the bond $b$ is chosen at 
random for a move in the $\rightarrow$ direction, the {\it a priori}
probabilities satisfy $p[{1\choose b}_p \to  {0\choose 0}_p] =
mp[{0\choose 0}_p\to {1\choose b}_p]$, since there are $m$ 
different 1-operators, but only one 0-operator, i.e., every accepted 
transition from any of the b 1-operators leads to the same 0-operator, 
but a transition from the 0-operator only leads to any 1-operator for a
given bond with probability $1/m$. One can easily verify that we 
satisfy the detailed balance condition by choosing
\begin{eqnarray}
\nonumber P\Biggl\lbrack {0\choose 0}_p \rightarrow {1\choose b}_p
\Biggr\rbrack &=&
{m\beta J\over 2(L-n)}\\
P\Biggl\lbrack {1\choose b}_p \rightarrow {0\choose 0}_p\Biggr\rbrack &=&
{2(L-n+1)\over Jm\beta},
\end{eqnarray}
where a number greater than 1 on the right hand side should be interpreted
as probability one.

The second kind of move is very similar, but it exchanges type-0
and type-2 operators. Again the string is sequentially
searched for these two kinds of operators, and detailed balance
is satisfied with the following exchange probabilities:
\begin{eqnarray}
\nonumber P\Biggl\lbrack {0\choose 0}_p \rightarrow {2\choose i}_p
\Biggr\rbrack &=&
{m\beta h\over (L-n)}\\
P\Biggl\lbrack {2\choose i}_p \rightarrow {0\choose 0}_p\Biggr\rbrack &=&
{(L-n+1)\over hm\beta}.
\end{eqnarray}

The third kind of move attempts to exchange type-1 and type-3 
operators. We can see that the weight function is unaffected by this
kind of move, and detailed balance is satisfied, but there is a 
restriction. 
An attempt to exchange ${1\choose b}_p$ for ${3\choose b}_p$ will result in 
a zero-weight function if the spins on the two sites that are connected to bond
$b$ point in opposite directions. Hence one needs to check the spin
configuration before attempting to exchange a 1-operator for
a 3-operator, while a 3-operator can always be exchanged for a
1-operator.

The fourth kind of move is slightly more complicated and involves
exchanging pairs of 2- and 4-operators. The reason why we have to
exchange pairs of operators is that the state of the system has
to return to its original state when it has been propagated by the
whole operator sequence; $|\alpha(0)\rangle=|\alpha(L)\rangle$, and if a spin
is flipped in an intermediate state it has to be flipped back
at a later time. Hence we attempt to make exchanges of the form 
${2\choose i}_p{2\choose i}_q\leftrightarrow {4\choose i}_p{4\choose i}_q$.
But we have to be careful, since this move results in the spin
at site $i$ being flipped in all states $|\alpha (p)\rangle,\ldots,|\alpha (q-1)\rangle$.
In case $p>q$, then  $|\alpha (0)\rangle$ will be affected, and we have to flip
the spin at site $i$ in the state $|\alpha(0)\rangle$ that we store in the
memory. Also, if there is an operator of type ${3\choose i}_r$
or ${5\choose i}_r$ with $p<r<q$, then this exchange would lead to
a zero weight function and the move must not be made. In order to
make an efficient up-dating scheme the operator sequence is divided
up into $n$ subsequences, each containing the necessary information
to make the above exchange for site $i \in \{1,\ldots n\}$.

It is easiest to actually store three lists: $T_i,P_i$ and $F_i$,
that contain information about the $i$th site. The list
$T_i=\{t_1,\ldots,t_{L(i)}\}$, with $t\in\{2,4\}$ contains all 2- and 
4-operators from the full operator string operating on site $i$.  
The list $P_i=\{p_1,\ldots,p_{L(i)}\}$,  contains
the positions $p$ of these operators in the full operator string.
The final list is $F_i=\{f_1,\ldots,f_{L(i)}\}$, where $f_j\in\{0,1\}$ 
indicates
if there are any operators of kind 3 or 5 between the positions
$p_j$ and $p_{j+1}$. The move then simply consists of choosing a
$t_j$ with the corresponding $f_j=0$. Thereafter one moves forward
in the $F$ list until the first nonzero $f_l$ is encountered. Then
one of the coordinates $k$ ($j<k<l$) is chosen and if $t_j=t_k$
the move is accepted, but if $t_j\ne t_k$ the move is rejected,
since the two operators then are of a different kind. If $t_j\ne t_k$
one can, however, permute the two operators. Again the
weight function in unaffected by this move, and the 
detailed-balance principle is automatically satisfied.

The fifth move involves another pair exchange of the form
${1\choose b}_p{1\choose b}_q\leftrightarrow {5\choose i}_p{5\choose i}_q$.
Again we have to be careful since this move flips the two 
spins connected to bond $b$ in all states $|\alpha (p)\rangle,\ldots,|\alpha(q-1)\rangle$,
and the weight function will vanish if there is an operator of type 
${3\choose b'}_r$ or ${5\choose b'}_r$, with $p<r<q$, where the bond $b'$ is 
connected to either of the sites that  bond $b$ is connected to. The full 
operator sequence is therefore again divided up into substrings containing the 
operators that act on a particular bond and the sites that it is connected to. 
One can make this kind of move in two different substrings independently of 
each other, as long as the two bonds do not connect, i.e. as long as they
do not have any site in common. The $n$ bonds on a one-dimensional
lattice can be partitioned in all the bonds that start at an odd
site, or all the bonds that start on an even site. Hence the
full operator string is split up two times, each time into
$n/2$ sub-strings, which can be updated independently. A two-dimensional
lattice has to be divided into four separate partitions; see 
Fig.~\ref{fig:lat}.

\begin{figure}
\centering
\epsfxsize=6cm
\leavevmode
\epsffile{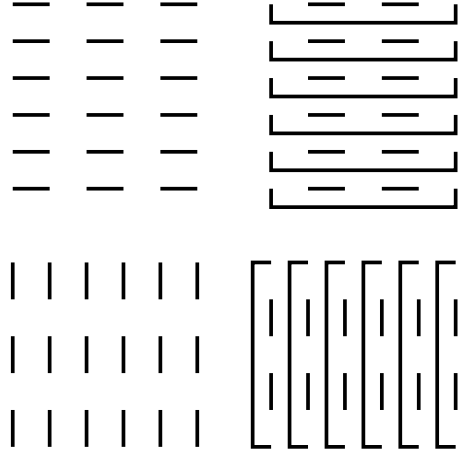}
\vskip0.4cm
\caption{The four different bond partitions of a $6\times 6$ lattice
with periodic boundary conditions.}
\label{fig:lat}
\end{figure}

The updating of the $n/2$ sublists is very similar to the updating
scheme for the fourth move. Four lists are stored this time.
The list $T_b=\{t_1,\ldots,t_{L(i)}\}$, with $t\in\{1,5\}$ contains all 1- 
and 5-operators from the full operator string operating on bond $b$.
The list $P_b=\{p_1,\ldots,p_{L(i)}\}$ again contains
the positions $p$ of these operators in the full operator string.
The list  $F_b=\{f_1,\ldots,f_{L(i)}\}$, where $f_j\in{0,1}$ indicates
if there are any operators of kind 3 or 5 operating on a bond connected
to bond $b$ between the positions $p_j$ and $p_{j+1}$. The final list
is $S_b=\{s_1,\ldots,s_{L(i)}\}$, where $s_i=1$ indicates that the spins 
connected to bond $b$ are aligned in state $|\alpha (p_i)\rangle$, and $s_i=0$ 
indicates that  they are anti-aligned. The move is identical to the procedure 
described for  the fourth move, except that now the move is canceled if 
$t_j\ne t_k$ or $s_j=1$.

The sixth move is the most complicated one, because it involves exchanges
between three different kinds of operators. It is this move that makes
this simulation particularly efficient, since it can generate configurations
of non-zero winding number. 

The following exchanges are attempted:
${4\choose i_1}_p{4\choose i_2}_q\leftrightarrow {1\choose b}_p
{5\choose b}_q$, where $i_1$ and $i_2$ are the two sites connected
to bond $b$. Again the exchange involves spin flips, and the lattice
is divided up into partitions as in move 5. Four lists are again used,
where $T_b$ includes all 1-,4- and 5-operators, $P_b$ gives the position
of the operators in the full sequence, $F_b$ indicates whether an
operator of type 3 acts on the two sites and $S_b$ keeps track of
the spin configuration. In the same fashion as above two operators,
$t_j$ and $t_k$, are chosen, and if $\{t_j,t_k\}=\{1,5\}$ an attempt
to change the two operators to $t_j=t_k=4$ is made, if the spin configuration
allows for this. If $t_j=t_k=4$ an attempt to change the two operators 
to $\{t_j,t_k\}=\{1,5\}$ is made. The detailed-balance principle
is satisfied if
\begin{eqnarray}
P\Biggl\lbrack {1\choose b}_p{5\choose b}_q \rightarrow 
  {4\choose i_1}_p{4\choose i_2}_q\Biggr\rbrack&=& 2h^2/J^2\\
P\Biggl\lbrack {4\choose i_1}_p{4\choose i_2}_q \rightarrow 
  {1\choose b}_p{5\choose b}_q\Biggr\rbrack&=& J^2/2h^2.
\end{eqnarray}

If the last move is excluded one can see that all possible
operator sequences are not sampled if periodic boundary conditions 
are used. To visualize this, we consider a four-site system for which is
is easy to realize that the simple string $H_{5,1}H_{5,2}H_{5,3}H_{5,4}$
cannot be reached. This is an example of a configuration with
a non-zero winding number (provided that the operator string operates
on an appropriate state). In order to make this configuration
possible we need to introduce a ``ring'' move. If we picture each
5-operator as connecting the two sites it acts on, we can see
that the configuration $H_{5,1}H_{5,2}H_{5,3}H_{5,4}$  creates a ring
around the system. The ring move is accomplished as follows:
starting at a random point in the operator sequence for a $n$-site
system a search for a set of $n/2$ different 5-operators is made.
If such a set is found, an attempt to exchange it with its
complementary set is made. An example for the 4-site system would be
$H_{5,1}H_{5,3} \to H_{5,2}H_{5,4}$. If the move is successful the winding
number has changed. Whether or not the move can be carried out of
course depends on constraints imposed by the states. As the system
size is increased beyond about $L=16$ it becomes virtually impossible
to perform ring moves.

In a one-dimensional system the ring move is the only move that
has to be added if the sixth move is not performed. In two dimensions
we can picture a ring around a small part of the system. We can, for
example, draw a ring around a $4\times 4$ square and we realize that we cannot
reach a configuration consisting of the 14 5-operators that connect
these sites. We therefore introduce a plaquette move. This move
changes the 5-operators that act on a single plaquette, which for
a two-dimensional square lattice is the smallest square in the lattice.
A plaquette move that involves arbitrarily large parts of the system,
for example the $4\times 4$  square considered above, can be reached with
these fundamental plaquette moves. The move is identical to the
one-dimensional ring move for a 4-site ring. 

In higher dimensions, when using periodic boundary conditions, we also
have to perform a direct generalization of the one-dimensional ring
move.  In a two-dimensional system with periodic boundary conditions
we can picture making rings around the system in both spatial
directions, and in three dimensions we could make the move in all
three spatial directions.  Apart from performing the move in all
spatial directions it is identical to the 1D ring move, and for a
linear system size larger than about $L=16$ ring moves are no longer
accepted. Note that this ring move can not be accomplished by the
above plaquette moves.

Without the external field in the $x$-direction, the Hamiltonian
conserves the total magnetization in the $z$-direction; $M_z = \sum_i S^z_i$,
and the operator string updates alone therefore sample only within a
fixed magnetization sector. In order to sample in the grand canonical 
ensemble, one then has to carry out an update that changes the
magnetization of the state $|\alpha \rangle$ in Eq.~(\ref{wn}).
A spin $S^z_i$ in $|\alpha \rangle$ can be flipped, without change in the
weight, if there are no operator $3 \choose b$ or $5 \choose b$ in
$S_L$ with $b$ a bond connected to spin $i$. The probability of this 
being the case approaches zero rapidly as $T$ is lowered, and hence
this update can be carried out only at relatively high temperatures. 
With an external field in the $x$ direction the magnetization is no
longer conserved and the simulation is automatically grand canonical.
It is nevertheless useful to carry out the single-spin flips at
high temperatures.

With the field present in two dimensions, the sixth move also makes 
plaquette and ring moves unnecessary. This can be understood since 
the sixth move introduces single 5-operators, and through a series 
of moves any configuration of 5-operators can be generated. Had we 
chosen the field in the $z$-direction this would not have been possible. 
Having the field in the $x$-direction therefore makes the simulation 
particularly efficient by introducing single-spin flipping operators. 
Strictly speaking we do not need to make the fifth move either, 
but especially at small fields it may be good to include this move 
to shorten the auto-correlation time and speed up the thermalization 
of the system.

Another advantage of choosing the $x$-direction for the field is that
we can then easily measure spin-spin correlations both parallel and
perpendicular to the field. The perpendicular correlation functions
are diagonal in our chosen basis, and can be evaluated according to
Eq.~(\ref{diaexp}). The parallel correlations involve expectation
values of products of the number of field operators in $S_L$ as
discussed in Refs.~\onlinecite{Sand}. We are particularly interested
in the perpendicular correlations for this model, since they are the
ones required to calculate the relaxation rate $1/T_1$. The
magnetization $M=\langle S^+_i+S^-_i\rangle /2$ is according to
Eq.~(\ref{exphm}) proportional to the average total number of field
operators $H_{4,i}$ in the sequence; $M=\langle n_4 \rangle /(N\beta
h)$.

A complete Monte Carlo step for the two-dimensional lattice, as used in 
this work, consists of
\begin{enumerate}
\item
Move 1, ${0\choose 0}_p\leftrightarrow {1\choose b}_p$, attempted at all
positions $p=1,\ldots L$ in the sequence $S_L$.
\item
Move 2, ${0\choose 0}_p\leftrightarrow {2\choose i}_p$, attempted at all
positions $p=1,\ldots L$ in $S_L$.
\item
Move 3, ${1\choose b}_p\leftrightarrow {3\choose b}_p$, attempted at all
positions $p=1,\ldots L$ in  $S_L$.
\item
The sequence is split up into $n$ sub-sequences and in each sequence
move 4,${2\choose i}_p{2\choose i}_q\leftrightarrow {4\choose i}_p,
{4\choose i}_q$ is attempted, whereafter the full operator sequence
is restored.
\item
The operator string is split up 4 times into $n/2$ different sub-sequences. 
In each sub-sequence the fifth move, 
${1\choose b}_p{1\choose b}_q\leftrightarrow {5\choose b}_p
{5\choose b}_q$, is then attempted a number of times of the order of the
length of the subsequence. The updated sub-sequences are
recombined into the full string.
\item
The operator string is split up 4 times into $n/2$ different 
sub-sequences and in each sub-sequence the sixth move, ${4\choose i_1}_p
{4\choose i_2}_q\leftrightarrow {1\choose b}_p {5\choose b}_q$, is 
attempted. The sub-sequences are recombined to the full string.
\item
All spins that are not acted on by any interaction operators $3\choose b$
or $5 \choose b$ are flipped with probability $1/2$.
\end{enumerate}

\begin{figure}[h]
\centering
\epsfxsize=7cm
\leavevmode
\epsffile{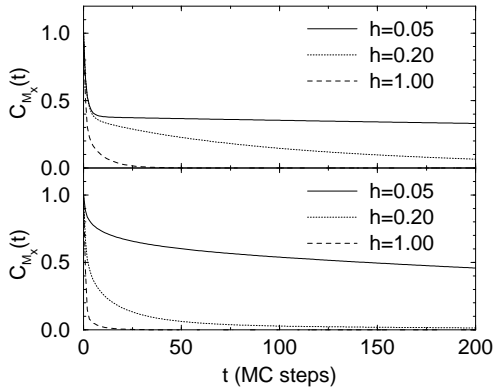}
\vskip0.4cm
\caption{Autocorrelation function for the $x$-component of the 
magnetization of  $8 \times 8$ lattices at three different strengths 
of the external field. The upper panel is for temperature $T/J=1$,
and the lower panel for $T/J=1/4$.}
\label{fig:amx}
\end{figure}

\begin{figure}[h]
\centering
\epsfxsize=7cm
\leavevmode
\epsffile{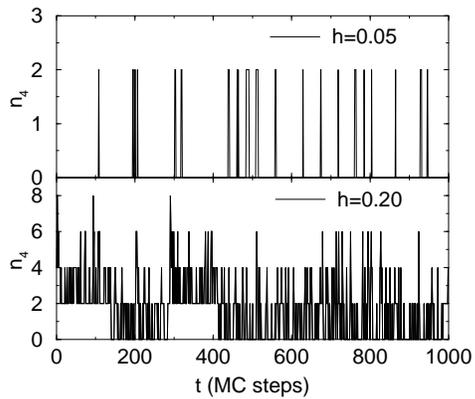}
\vskip0.4cm
\caption{The time dependence of the number of field operators in the
operator string at $T/J=1$ and two different field strengths.}
\label{fig:tmx}
\end{figure}

In principle it is possible to combine moves 1-3 into a single move
with several ``branches''. For simplicity we have not discussed this
(only marginally) more complicated approach here.

A full simulation consists of a number of equilibration steps, during
which the maximum length of the string is automatically increased as
the string grows. The length of the string very quickly reaches its
equilibrium, and after the equilibration steps are done the maximum
length is fixed (as described in the previous section) and
measurements are carried out at even intervals.

The present updating scheme is very efficient provided that the density 
of single-spin flipping operators $H_{4,i}$ in the sequence is not too 
small. In practice, we have found that simulations for $h /J \agt 0.05$ 
deliver accurate results without too much effort, independently of $T$. 
In Figure \ref{fig:amx} we show some results for the auto-correlation 
function of the magnetization, defined according to
\begin{equation}
C_{M_x}(t) = {\langle M_x(i)M_x(i+t)\rangle - \langle M_x\rangle^2
\over \langle M_x^2\rangle - \langle M_x\rangle^2 },
\end{equation}
where $M_x(i)$ is the value of the magnetization estimator, $n_4 /(N\beta h)$,
at the $i$th Monte Carlo step. We show results for $8\times 8$ lattices 
at two different temperatures and three different field strengths. Note 
that the long-time correlations decrease with decreasing temperature. 
This is related to the fact that the distribution of magnetization 
values becomes broader. For example, for $h/J=0.05$ at $T/J=1$, the 
magnetization essentially fluctuates between two values, corresponding 
to configurations with $0$ or $2$ field operators, as shown in Figure 
\ref{fig:tmx}. The system spends long times in states with no field 
operators and occasionally short times with 2 operators. These two 
time scales are reflected in the autocorrelation function, which shows 
a rapid decay at short times, but a very slow decay at longer times 
(the asymptotic exponential decay time is several hundred Monte Carlo 
steps). At the lower temperature the short-time behavior shows a less 
rapid decay, but also a faster asymptotic decay, as the fluctuations 
in the distribution of the number of field operators $n_4$ is now broader, 
and the likelihood of a fluctuation by $\pm 2$ in the simulation is higher.
At $h/J=0.2$, the time dependence of $n_4$ shows fluctuations on much
shorter time-scales, and the autocorrelation function accordingly
decays considerably faster.

\section{THE MAXIMUM ENTROPY TECHNIQUE}
It is notoriously difficult to obtain dynamic information from
quantum Monte Carlo simulations. One of the presently most common and
promising methods is the maximum entropy technique,~\cite{Jar} which 
we used to calculate the spin-lattice relaxation rate.  In this 
section we will discuss various ways to use the method, compare 
results to exact diagonalization and show some of the limitations 
of the procedures.
 
The nuclear magnetic resonance spin-lattice relaxation rate is given by
~\cite{Slic}
\begin{equation}
{1\over T_1}={1\over N}\sum_q |A_q|^2S(q,\omega_N),
\end{equation}
where $A_q$ is the Fourier transform of the hyperfine coupling.
The resonance frequency $\omega_N$ is so small compared to
other energy scales that we consider the limit 
$\omega\rightarrow 0$.
From now on we assume that $A_q=1$. The dynamic structure factor 
$S(q,\omega)$, which measures transverse spin correlations, 
can be obtained from the imaginary time dependent correlation function
\begin{equation}
C(q,\tau)=\sum_r e^{iqr}C(r,\tau)=\sum_r e^{iqr}\Bigl\langle 
S^z_i(\tau)S^z_{i+r}(0)\Bigr\rangle
\end{equation}
by inverting the expression
\begin{equation}
\sum_qC(q,\tau)=\int d\omega\sum_qS(q,\omega)e^{-\tau\omega}=C(r=0,\tau).
\end{equation}
Calculating the local imaginary time correlation function is
therefore in principle sufficient for determining the relaxation rate, 
after performing a maximum entropy inversion. For reasons that will 
become clear below, we did, however, also measure the spatial dependence 
of the correlation function.

Using conservation of momentum it is possible to obtain 
finite temperature exact diagonalization results for system 
sizes up to $4 \times 4$. Below we will show diagonalization data
compared to Monte Carlo + maximum entropy results.

First a few words concerning the default model needed in the maximum
entropy method. All data shown in this paper have been generated using a
flat default model as defining zero entropy. Much has been written about
how to choose a good default model,~\cite{Jar} but from our experience 
it seems that unless something very specific and very accurate
is known about the solution (in which case it probably is
unnecessary to use Monte Carlo + maximum entropy),  one should 
use a flat default model. If only general features, obtained from,
for example, perturbation theory or some analytical mean field
solution, are known about the function, then one introduces a bias 
towards one approximate model by placing peaks at frequencies that are only 
approximate. If the answer is very dependent on what default model 
one uses, then the accuracy of the results should be viewed with caution.
Using good data and a flat default model is optimal in the
sense that no prior bias has been used to  construct features
in the spectral function; the default model only has a 
regularizing effect. We worked extensively with
Gaussian default models, where we determined all the parameters
of the default model through sum rules that could be calculated
in the Monte Carlo simulation. In some cases this worked fairly well,
but in those cases a flat default model worked almost equally
well. In other cases the maximum entropy solution would be very
close to the Gaussian default model, but far from the exact
diagonalization results. This clearly shows that the problem
is ill-posed. In these instances it was our experience
that a flat default model resulted in better agreement with exact 
diagonalization.

\begin{figure}
\centering
\epsfxsize=8cm
\leavevmode
\epsffile{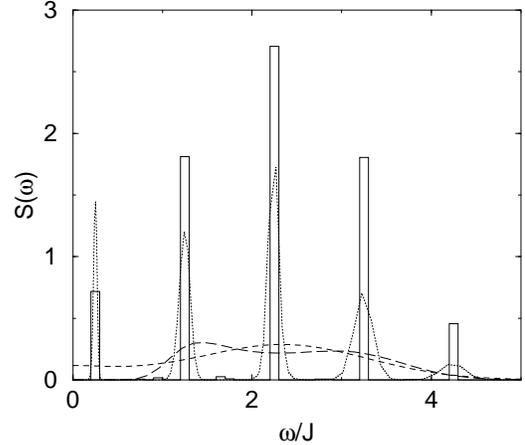}
\caption{Maximum entropy compared to exact diagonalization for a
$4 \times 4$ system at $T/J=0.25$ and $h/J=0.25$. The exact results are 
represented by the solid-line histogram. The maximum entropy method is 
applied to each momentum separately (dotted curve), to the average of 
all momenta (dashed curve) and to the average of all momenta, except 
the $q=(0,0)$ (long-dashed curve).}
\label{fig:t025}
\end{figure}

We now turn to actual comparisons with exact diagonalization results.
At low temperatures the spectral function is dominated by a delta
function for each magnon, and this extreme limit will first be
considered to illustrate some weaknesses and strengths of the maximum
entropy solution. In Fig.~\ref{fig:t025} we show results for a $4
\times 4$ system at inverse temperature $T/J=0.25$ and field strength
$h/J=0.25$. As usual, the exact diagonalization result really only
consists of a series of delta peaks, but the spectrum has been divided
into bins of finite width in order to illustrate the result more
clearly. The first peak is at exactly $\omega=h/J=0.25$ and is the
response of the $q=(0,0)$ momentum. This peak will remain a delta
function even at finite temperatures since the
spin-correlation function then will contain the {\it total} spin raising and
lowering operators, and hence transitions can occur only between levels
within the same spin multiplet. These levels are all separated by the
Zeeman splitting, which is independent of temperature. Therefore the
delta peak at the Zeeman energy will not get broadened by
temperature. For finite q-values there will be transitions between
different spin multiplets causing transition energies that differ from
the Zeeman energy.

In order to be able to directly study the different momenta we have
measured the full spatial dependence of the correlation function and
we can therefore either analytically continue each momentum and then
average the spectrum, or first average the correlation function and
then do the analytic continuation.  In Fig.~\ref{fig:t025} both
methods are demonstrated. The dotted curve shows the result when
analytically continuing each momentum separately. We notice two
important features.  First, the maximum entropy method resolves each
peak, and secondly, the resolution of the peaks decreases with
frequency, as can be expected, since the factor $e^{-\tau\omega}$
makes the inversion insensitive to high frequency features. The dashed
curve shows the result when the average over all momenta is continued.
We notice immediately that the maximum entropy method has difficulty
in resolving more than one peak and tends to smear the result
out. Focusing on the limit $\omega\rightarrow 0$ we notice that
because of the $q=(0,0)$ peak at $\omega=h/J=0.25$, an estimate of
$1/T_1$ would be too high, since the weight of the $q=(0,0)$ peak is
smeared out all the way to $\omega=0$. To remedy this effect we can
average over all momenta, except $q=(0,0)$. The results are shown in
the dot-dashed curve, and we see that this certainly affects the
solution close to $\omega=0$, where there is now correctly, no weight.

\begin{figure}
\centering
\epsfxsize=8cm
\leavevmode
\epsffile{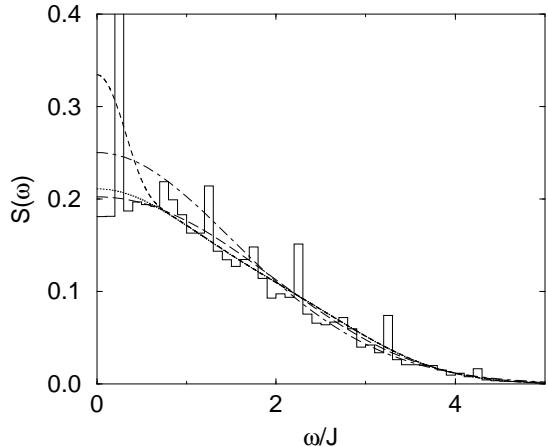}
\caption{Maximum entropy results compared to exact diagonalization data
for a $4\times 4$ system at $T/J=2.0$ and $h/J=0.25$. 
The exact results are represented by the solid-line histogram. 
The maximum entropy method is applied to each
momentum separately but not including $q=(0,0)$ (dotted curve), 
to each momentum separately including $(0,0)$ (dashed curve),
to the average of all momenta not including $(0,0)$ (long-dashed
curve) and to the average of all momenta including $(0,0)$ (dot-dashed 
curve).}
\label{fig:t200}
\end{figure}

We are, however, interested in  results for large system sizes,
for which the spectral function should be smooth on a reasonably fine
frequency scale. We cannot compare our results to diagonalization
results for more than $4 \times 4$ sites. We can, however, study this
system at higher temperatures, for which the spectral function is 
fairly smooth. In Fig.~\ref{fig:t200}
the same parameter values as in the previous figure are shown, but
at a higher temperature, $T/J=2.0$. We see that the diagonalization
results are much smoother and we can analyze the maximum entropy results.
The dashed curve shows all momenta separately continued. $1/T_1$
is grossly overestimated since the $q=(0,0)$ peak is smeared
out at this higher temperature. Removing the $q=(0,0)$ peak gives
a much better estimate; see the dotted curve. Continuing the average
of all momenta again places too much weight at low frequencies,
because of the $q=(0,0)$ peak, as can be seen from the dot dashed
curve. Removing the $q=(0,0)$ momentum from
the average improves the result a great deal; see the long dashed
curve. 

We found that the optimal use of the maximum entropy method in our
case was to continue the average of all momenta, excluding the
$q=(0,0)$ momentum. For this momentum  we know the exact analytic result, 
and including it in the continuation  leads to an over-estimate
of $1/T_1$. The reason for not continuing each separate momentum,
which worked miraculously well in Fig.~\ref{fig:t025}, is that
at intermediate temperatures this method also leads to an
overestimation of $1/T_1$, since peaks close to the origin
will be smeared out, and much of their weight will be incorrectly
placed  at $\omega=0$.

To conclude this Section, we have found that the maximum entropy
method is a useful method to obtain real time data from imaginary
time quantum Monte Carlo data for the Heisenberg ferromagnet.
We believe that the specific recipe for how to use the method 
probably varies from system to system, but for the case we have considered
here we found the most useful default model to be flat, and we also 
found that we had to separate out the $q=(0,0)$ momentum before 
carrying out the analytic continuation.

Similar calculations of NMR relaxation rates have previously been
carried out also for various quantum antiferromagnets.~\cite{Sand3} In
that case the differences between maximum entropy calculations based
on momentum and real-space correlations were less pronounced than what
we have found here. The ferromagnetic case appears to be more
difficult since the field induces additional structure in the
frequency dependence.  Also, at low temperatures independent magnons
are exact excitations of the ferromagnet, which causes further sharp
features in the spectral function.

\section{RESULTS}
\subsection{Magnetization}

The Heisenberg model is one of the basic non-trivial models of a quantum 
ferromagnet. A detailed knowledge of the field dependence of the 
magnetization is therefore of interest for comparisons both with
analytical and experimental results. In this Section we will show the 
general field dependence of the magnetization for the two-dimensional
model. We will also compare our Monte Carlo results to exact diagonalization
results and show the nature of the finite size corrections. We also 
compare the results to recent experimental results for the $\nu=1$ 
quantum Hall state. 

In Fig.~\ref{fig:finites} we show a comparison of exact diagonalization
and Monte Carlo data for a $4 \times 4$ system. We have verified the agreement 
to a relative error of $10^{-4}$. To this degree of accuracy the results for
the largest system sizes shown ($16 \times 16$ and $32\times 32$) 
have also converged. We did additional test runs for systems of 
size $64\times 64$, and the results were within error bars of the 
above results. From Fig.~\ref{fig:finites} we notice that
the finite size effects are largest at intermediate temperatures,
which is easily understood since in the limit of very low temperatures
the magnetization will be fully saturated independently of the
lattice size, while in the high temperature limit the magnetization
will vanish independently of the system size. 

\begin{figure}[h]
\centering
\epsfxsize=8cm
\leavevmode
\epsffile{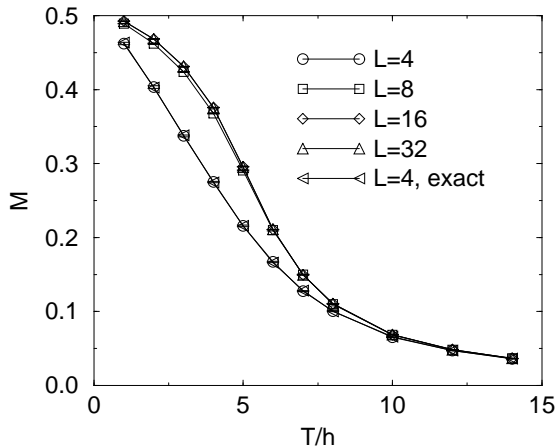}
\caption{Monte Carlo results for the magnetization vs temperature for 
$L \times L$ systems with $L=4,8,16$ and $32$, and exact
diagonalization results for $L=4$. The magnetic field $h=0.1$.}
\label{fig:finites}
\end{figure}

\begin{figure}[h]
\centering
\epsfxsize=8cm
\leavevmode
\epsffile{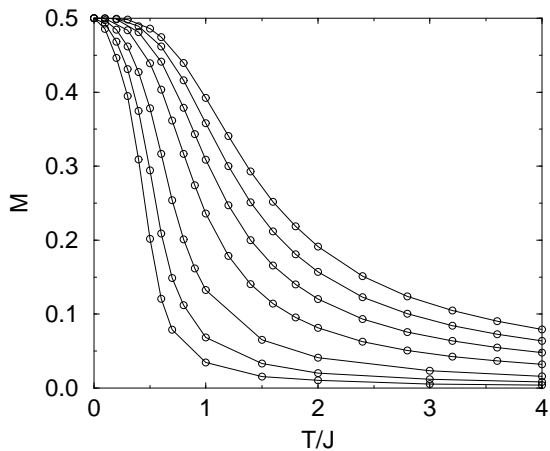}
\caption{Magnetization vs temperature for the 2D Heisenberg ferromagnet
in magnetic fields of strengths h=1.00, 0.80, 0.60, 0.40, 0.20,0.10 and
0.05, from top to bottom. The results were calculated using lattices 
sufficiently large  (typically $32 \times 32$) to eliminate finite-size 
effects almost completely.}
\label{fig:mag}
\end{figure}

In Fig.~\ref{fig:mag} we present a plot of the magnetization as a function 
of temperature for a range of field strengths. We have tried to scale
the data as a function of $T/h^\alpha$ for different values of the exponent 
$\alpha$. Such scaling does not collapse the data onto a single curve, 
however. Analytical calculations have also suggested that there is no scaling
in $T$ and $h$ for this model.~\cite{Read}

\begin{figure}
\centering
\epsfxsize=8cm
\leavevmode
\epsffile{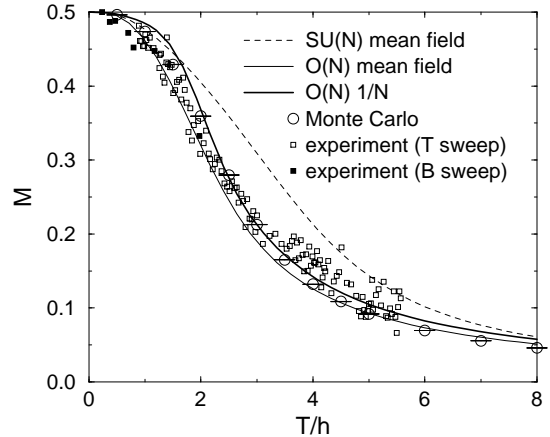}
\caption{Magnetization curve for the $\nu=1$ quantum Hall ferromagnet. 
Comparison of Schwinger boson [SU({\it N}) and O({\it N}) symmetric versions of
the theory], quantum Monte Carlo and experimental data (Ref. 2). The 
calculations were carried out at field strength $h/J=0.32$.
}
\label{fig:exp}
\end{figure}

We also show a comparison of recent magnetoabsorption
measurements~\cite{Manf} performed on the $\nu=1$ quantum Hall state
compared to quantum Monte Carlo data and Schwinger boson calculations
in Fig.~\ref{fig:exp}. There are no adjustable parameters in this
comparison, since $J$ can be calculated exactly for the zero-width
well~\cite{Moon} using the exactly known spin wave dispersion 
for the quantum Hall ferromagnet, and estimated for the actual experimental
system~\cite{Macd}. The excellent agreement shows that this itinerant
ferromagnet is well described by an effective Heisenberg model.  As
discussed in our previous paper~\cite{Timm} the comparison with the
analytical Schwinger boson solutions showed that the $1/N$ corrections
to the O$(N)$ model agrees well with Monte Carlo and experimental
results at moderate and intermediate temperatures, while the
low-temperature result was better reproduced by the SU$(N)$ model. For
a more comprehensive discussion of the Schwinger boson results and the
experimental data, we refer to our previous paper.~\cite{Timm}

\subsection{Spin-lattice relaxation rate}
In Fig.~\ref{fig:relax} we compare mean-field results~\cite{Read,Timm}
for the nuclear magnetic relaxation rate $1/T_1$, to numerical results
obtained by analytic continuation of imaginary time Monte Carlo data,
as described previously. As we have discussed, the maximum entropy
results have to be viewed with some caution. The error bars are
obtained using the bootstrap technique.~\cite{Num} They do not
strictly correspond to a statistical likelihood that the estimated
relaxation rate is within error bars of the true relaxation rate,
since there is also an unknown systematic error due to the bias of the
maximum entropy procedure.  The error bars do contain information
about how sensitive the results are to the variations in the MC data.

We notice a fairly good agreement between the analytic and numerical
results and, as in the case for the magnetization,~\cite{Timm} the
O($N$) theory appears to be somewhat closer to the numerical results.
Notice that this time the numerical results do not lie between the
O($N$) and SU($N$) solutions, which was the case for the
magnetization. At zero temperature the relaxation rate is zero, since
no spins flip in the ordered ferromagnet and therefore the nuclear
spins cannot relax. At low temperatures the rate is activated and
caused by thermally activated spin waves.

Not enough experimental data is available for the relaxation rate to make a 
comparison with the above numerical results.
Because results for the Heisenberg  model agreed very well with experimentally
measured magnetization, it would be of interest to 
compare the relaxation rate, to see if the Heisenberg  model also
captures the correct dynamic behavior of the $\nu =1$ quantum Hall 
ferromagnet.

\begin{figure}
\centering
\epsfxsize=8cm
\leavevmode
\epsffile{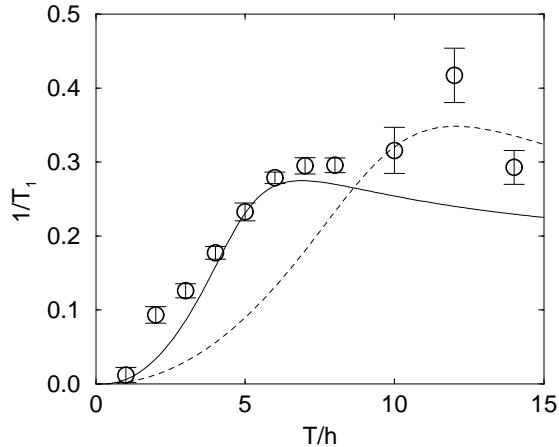}
\caption{Monte Carlo and maximum entropy numerical results (circles)
and Schwinger boson results [SU(N) mean field (dashed curve) and O(N) mean 
field (solid curve)] for the relaxation rate $1/T_1$ 
for $h/J=0.10$.}
\label{fig:relax}
\end{figure}

\section{SUMMARY AND CONCLUSION}

We have described an approximation-free quantum Monte Carlo technique
and applied it to a two-dimensional ferromagnet in a magnetic field.
We have shown that applying the field in the transverse direction
causes the simulation to become free of systematic errors although
only local Monte Carlo moves are made. The transverse field also
enables easy access to transverse spin correlation functions.  We have
calculated the temperature dependence of the magnetization for a range
of field strengths and discussed finite-size effects.  Some results
have previously been compared to measurements of the $\nu=1$ quantum
Hall state and analytical Schwinger Boson calculations.~\cite{Timm}
The high relative accuracy ($10^{-4}$) of the Monte Carlo results made
it possible to make statements about the relative merits of the
SU($N$) and O($N$) solutions. We have, in addition, calculated the
spin-lattice relaxation rate $1/T_1$ using a maximum entropy method,
and compared also these results with the Schwinger boson results.  The
role of the maximum entropy default model was discussed, and for the
2D Heisenberg model we argued that a flat default model works best. We
further discussed advantages and problems arising when applying the
maximum entropy procedure to imaginary time correlation functions in
real space and momentum space.

\section{ACKNOWLEDGMENTS}

The research was supported by NSF Grants No. DMR-9714055, CDA-9601632,
DMR-9629987, and DMR-9712765. P.H. acknowledges support by Suomalainen 
Tiedeakatemia and C.T. by the Deutsche Forschungsgemeinschaft.

\end{document}